# Graphene on hexagonal boron nitride as a tunable hyperbolic metamaterial

S. Dai[1], Q. Ma[2], M. K. Liu[1,3], T. Andersen[2], Z. Fei[1], M. D. Goldflam[1], M. Wagner[1], K. Watanabe[4], T. Taniguchi[4], M. Thiemens[5], F. Keilmann[6], G. C. A. M. Janssen[7], S. -E. Zhu[7], P. Jarillo-Herrero[2], M. M. Fogler[1], D. N. Basov[1]*.

[1]*Department of Physics, University of California, San Diego, La Jolla, California 92093, USA*

[2]*Department of Physics, Massachusetts Institute of Technology, Cambridge, Massachusetts 02215, USA*

[3]*Department of Physics, Stony Brook University, Stony Brook, New York 11794, USA*

[4]*National Institute for Materials Science, Namiki 1-1, Tsukuba, Ibaraki 305-0044, Japan*

[5]*Department of Chemistry and Biochemistry, University of California, San Diego, La Jolla, California 92093, USA*

[6]*Ludwig-Maximilians-Universität and Center for Nanoscience, 80539 München, Germany*

[7]*Micro and Nano Engineering Lab, Department of Precision and Microsystems Engineering, TU Delft, Mekelweg 2, 2628 CD Delft, The Netherlands*

*Correspondence to: dbasov@physics.ucsd.edu

**Hexagonal boron nitride (h-BN) is a natural hyperbolic material[1], for which the dielectric constants are the same in the basal plane ($\varepsilon^t \equiv \varepsilon^x = \varepsilon^y$) but have opposite signs ($\varepsilon^t \varepsilon^z < 0$) from that in the normal plane ($\varepsilon^z$)[1-4]. Due to this property, finite-thickness slabs of h-BN act as multimode waveguides for propagation of hyperbolic phonon polaritons[1,2,5] – collective modes that originate from the coupling between photons and electric dipoles[6] in phonons. However, control of these hyperbolic phonon polaritons modes has remained challenging, mostly because their electrodynamic properties are dictated by the crystal lattice of h-BN[1,2,7]. Here we show by direct nano-infrared imaging that these hyperbolic polaritons can be effectively modulated in a van der Waals heterostructure[8] composed of monolayer graphene on h-BN. Tunability originates from the hybridization of surface plasmon polaritons in graphene[9-13] with hyperbolic phonon polaritons in h-BN[1,2], so that the eigenmodes of the graphene/h-BN heterostructure are hyperbolic plasmon-phonon polaritons. Remarkably, the hyperbolic plasmon-phonon polaritons in graphene/h-BN suffer little from ohmic losses, making their propagation length 1.5-2.0 times greater than that of hyperbolic phonon polaritons in h-BN. The hyperbolic plasmon-phonon polaritons possess the combined virtues of surface plasmon polaritons in graphene and hyperbolic phonon polaritons in h-BN. Therefore, graphene/h-BN structures can be classified as electromagnetic metamaterials[14] since the resulting properties of these devices are not present in its constituent elements alone.**

Van der Waals (vdW) heterostructures assembled from (one or few) monolayers of graphene, h-BN, $MoS_2$ and other atomic crystals in various combinations are emerging as a new paradigm to attain the desired electronic[8,15] and optical[16] properties. These heterostructures are also of interest in the context of polaritons that are ubiquitous in metals, insulators and semiconductors[6,16]. At least two different classes of propagating polaritons are firmly established in vdW systems: surface plasmon polaritons ($SP^2$) in graphene[9-13] and hyperbolic phonon polaritons ($HP^2$) in h-BN[1,2]. In graphene/h-BN meta-structures, coherent oscillations of the electron density in graphene and the atomic vibrations in h-BN produce hybridized plasmon-phonon modes. Surface plasmon-phonon modes[17] and related energy transfer processes[18] have been investigated in structures comprised of graphene with monolayer h-BN or a BN nanotube. However, neither monolayers[17] nor nanotubes[18] of BN support hyperbolic response: an exquisite attribute of three-dimensional specimens of this layered anisotropic material[1-4]. A remarkable feature of graphene/h-BN heterostructures uncovered in our experiments is that monolayer graphene impacts the hyperbolic response of h-BN slabs as thick as 99 nm, exceeding 300 atomic layers. We demonstrate that both the wavelength and intensity of hyperbolic polaritons can be controlled via electrostatic gating of the top graphene layer.

Direct experimental access to the tunable hyperbolic response in graphene/h-BN is provided by IR nano-spectroscopy and nano-imaging via a scattering-type scanning near-field optical microscope (s-SNOM) as shown in Fig. 1a (see also Methods). The same technique was utilized in a recent study[19] of h-BN/graphene/h-BN vdW heterostructures; however, the hyperbolic spectral regions were not probed therein. In Fig. 1b we show broad-band nano-IR spectra of the normalized (Methods) scattering amplitude $s(\omega)$ as a function of frequency $\omega = 1 / \lambda_{IR}$, $\lambda_{IR}$ being the IR wavelength, for h-BN, SiO$_2$ substrate, and graphene/h-BN meta-structures. The spectra for SiO$_2$ (black) and h-BN (red) display resonances due to their mid-IR phonons[1,20]. The two hyperbolic regions of h-BN[1,2] are highlighted in Fig. 1b: the type I region where $\varepsilon^z < 0$, $\varepsilon^t > 0$, extends over the frequency range $\omega = 746 - 819$ cm$^{-1}$, and type II region where $\varepsilon^z > 0$, $\varepsilon^t < 0$, spans the range $\omega = 1370 - 1610$ cm$^{-1}$. Both type I and II resonances of h-BN are modified in meta-structures incorporating monolayer graphene (the blue spectrum in Fig. 1b). The impact of graphene is particularly prominent in the type I region where the resonance mode is significantly enhanced and blue-shifted by nearly ~ 25 cm$^{-1}$ compared to the response of a standalone h-BN slab.

The peculiar electrodynamic response of graphene/h-BN is vividly illustrated by the calculated frequency ($\omega$) – momentum ($q$) dispersion relations of its polariton modes (Figs. 1c-e, see Supplementary Section 1 for details). Following Ref. 6, we visualize these dispersions using a false colour map of the imaginary part of the reflectivity $r_p$. It is instructive to first consider the polaritons of the two constituent elements (graphene and h-BN) separately. In Fig. 1c we plot the dispersion of SP$^2$ for a freestanding graphene layer for three selected values of the Fermi energy $E_F$. These parabolic curves are described by the equation[21] $q_p(\omega) = \frac{(\hbar\omega)^2}{2e^2 E_F}$. The corresponding plasmon wavelength is

$$\lambda_p = \frac{2\pi}{q_p} = \frac{4\pi e^2 E_F}{(\hbar\omega)^2}. \qquad (1)$$

Figure 1d plots the dispersion of HP$^2$s in an h-BN slab of thickness $d = 58$ nm on SiO$_2$ (no graphene). In a stark contrast to isotropic crystals, where longitudinal optical phonons occur at a single degenerate frequency $\omega_{LO}$, in h-BN, multiple distinct branches of HP$^2$ exist[1,2,5]. These different branches correspond to quantized HP$^2$ waveguide modes[1,2,5] with a scalar potential oscillating across the slab and having different number of nodes[22].

Each waveguide mode disperses between $\omega_{TO}$ and $\omega_{LO}$ (Fig. 1b). Our theoretical results and discussion below are relevant for all these modes. The experimental results mainly concern the principal mode, the nodeless waveguide mode of the lowest momentum. Finally, in Fig. 1e we display the dispersion of the new collective modes – hyperbolic plasmon-phonon polaritons ($HP^3$) – that arise from mixing of the $SP^2$ and $HP^2$ in the graphene/h-BN meta-structure. The graphene Fermi energy $E_F = 0.37$ eV was estimated from the surface polariton wavelength in Fig. 3d (see also Ref. 19). The modification of hyperbolic response by graphene is clearly manifested in the blueshift of the $HP^3$ frequencies with respect to those of $HP^2$ (Figs. 1d,e). The shift of momenta (at a fixed frequency) is opposite in the two hyperbolic bands: negative in the Type II band and positive in the Type I band (Supplementary Section 1 and 2). This contrasting behavior stems from the fact that the polariton dispersion being negative and positive in the Type I and II region, respectively.

The change of the polariton wavelength induced by graphene is described by the formula (Supplementary Section 1):

$$\Delta\lambda(\%) = \frac{\lambda_{HP3} - \lambda_{HP2}}{\lambda_{HP2}} \simeq \frac{\lambda_p}{\pi d} \frac{\varepsilon^z}{1 - \varepsilon^z \varepsilon^t} \qquad (2)$$

In a typical situation where $\tilde{\varepsilon}^z$, $\varepsilon^t$ are neither too large nor too small, this formula predicts that $\Delta\lambda(\%)$ is of the order of the ratio of the two length scales: the plasmon wavelength $\lambda_p$ of graphene and the thickness $d$ of h-BN. This clarifies why the influence of graphene remains substantial in h-BN as thick as $d = 300$ nm (value obtained from calculations in Supplementary Section 1). The length scale over which graphene can exert its influence on the electrodynamics of surrounding media is set by its plasmon wavelength.

Importantly, the plasmon wavelength can be controlled over a wide range through an applied gate voltage. Thus, HP$^3$s inherit the hyperbolic nature of HP$^2$s while gaining an important added virtue: tunability with applied gate voltage. Outside the two HP$^3$ regions, the plasmonic character of the dispersion is largely preserved (Fig. 1e). The polaritonic mode flattens out in the vicinity of $\omega_{TO}$ of either of the two hyperbolic bands: a consequence of mode repulsion[23]. Similar interactions between plasmons and phonons have been studied in graphene on other substrates (e.g., SiO$_2$, SiC, ion gel et al.) and monolayer h-BN[9-13,17,20,23,24], where the hyperbolic response is not supported. Following the terminology established there, we refer to the collective modes existing outside the h-BN hyperbolic bands as surface plasmon-phonon polaritons (SP$^3$)[17,19].

Infrared nano-imaging data (Figs. 2,3) visualizing the propagating polaritons in our meta-structures unambiguously support the above theoretical predictions. The basic principles of polariton imaging have been detailed elsewhere[1,5,10,11]. In short, when illuminated by the IR beam, the s-SNOM tip launches radially propagating polariton waves (Fig. 1a). The tip then registers the interference pattern between launched and edge-reflected polaritons, yielding oscillating fringes in the scattered near-field signal. The periodicity of the fringes is one-half of the polariton wavelength (denoted generically by $\lambda$, with suitable subscripts when needed).

In Fig. 2a we present nano-imaging data at a representative frequency $\omega = 1495$ cm$^{-1}$ for a meta-structure that includes a slab of h-BN (thickness $d = 25$ nm) partially covered by a heavily doped monolayer graphene. We observe polariton fringes in both covered (graphene/h-BN) and uncovered (h-BN) areas. In the uncovered h-BN region (the bottom half of Fig. 2a), the fringes originate from the Type II hyperbolic polaritons[1,5]. In the graphene/h-BN region (the upper part of the image in the middle of Fig. 2a) we observe fringes that are stronger and have a longer oscillation period. Prominent fringes can also be detected along the graphene edge (the dashed green line). Line profiles obtained normal to the h-BN edge (Fig. 2b) help to quantify the nearly 50% increase of both amplitude and wavelength of the fringe oscillations due to the presence of doped graphene. This prominent modification is attributed to plasmon-phonon coupling and the formation of the type II $HP^3$ band in our meta-structure (Figs. 1e and 2c).

We observe similar enhancement of polaritonic oscillations (Figs. 2a-b) at all $\omega$ within the Type II band. The blue dots in Fig. 2c display these data in the dispersion relation: $\omega$ plotted versus the polariton momentum $q$ that can be read off the line profiles as $q = 2\pi / \lambda$. For comparison, we also measured the $HP^2$ dispersion for pristine h-BN (red triangles in Fig. 2c). Both data sets match the theoretical calculations (false colour and white lines, Supplementary Section 1) for the principle branch of hyperbolic polaritons. In addition to the principal mode, polaritons from higher order branches are also enhanced in graphene/h-BN (Supplementary Section 3). The largest experimentally observed $\Delta\lambda(\%) = 90\%$ in this data set is reached at $\omega = 1545$ cm$^{-1}$. In comparison, the approximate equation (2) yields 98%, using $d = 25$ nm, $\lambda_p = 180$ nm (equation (1)), $\varepsilon^z = 2.77$, and $\varepsilon^t = -1.98$. The agreement between the experiment, analytical theory, and numerical simulations attests to the validity of the plasmon-phonon coupling approach to account for the modified spectrum of hyperbolic modes. We performed measurements for a variety of samples, e.g., graphene on h-BN of different thicknesses, graphene obtained by exfoliation and chemical vapor deposition (CVD) techniques, all of which produced consistent results.

The tuning of polaritons in the Type II HP$^3$ region via electrostatic gating (Methods) is presented in Figs. 3a-b at another representative frequency $\omega = 1395$ cm$^{-1}$. When graphene is close to charge neutrality (Fig. 3a), the profile of propagating polariton in graphene/h-BN is nearly indistinguishable from that of uncovered h-BN. Once graphene is doped by gating (Fig. 3b), both the intensity and wavelength of the polaritonic features were significantly increased. This systematic study of the gate-tunability is summarised in Fig. 3c (blue dots), where the wavelength consistently increases with the absolute value of gate voltage at fixed frequency $\omega = 1395$ cm$^{-1}$.

Here we stress the distinction between the electrodynamics in HP$^3$ and SP$^3$ spectral regions (Fig. 1e). The latter are localized on the sample surface whereas the former propagate through the entire graphene/h-BN meta-structure (Fig. 3d, inset) in the form of guided waves. We verified the waveguiding character by examining the thickness-dependence of HP$^3$ wavelength using multiple h-BN slabs covered by a large sheet of CVD graphene. The Fermi energy for all the graphene/h-BN samples was about the same, $E_F = 0.37$ eV. Both in experiment (blue dots) and simulations (green line), the dependence of the HP$^3$ wavelength $\lambda_{HP^3}$ on $d$ is nearly linear with a finite intercept (Fig. 3d and Supplementary Section 1), where $\Delta\lambda(\%)$ ranges from 70% ($d = 25$ nm) to 18% ($d = 99$ nm). This law readily follows from two analytical results: $\Delta\lambda(\%) \sim d^{-1}$ (equation (2)) and $\lambda_{HP^2} \sim d$ (Ref. 1). In contrast, the localized SP$^3$ modes show essentially thickness-independent behavior of the polariton fringes outside the hyperbolic region (e.g., $\omega = 882$ and 1617 cm$^{-1}$). The fundamental difference between HP$^3$ and SP$^3$ is further illustrated by polariton field simulations (yellow traces in Fig. 3d, inset). The field distribution of HP$^3$ in graphene/h-BN is characteristic of a standing wave, whereas that of the SP$^3$ is localized at the graphene/h-BN interface and decays evanescently in the interior of the h-BN.

We conclude by pointing out that tunable hyperbolic response in graphene/h-BN devices does not introduce evident losses (Fig. 2b). The loss factor of HP$^3$, defined as $\kappa/q$ for the complex momentum $q + i\kappa^6$, can be as small as 0.06 but increases up to ~ 0.10 in the vicinity of the longitudinal phonon mode. In fact, the propagation length of HP$^3$ in graphene/h-BN is factor of 1.5–2.0 longer than HP$^2$ in h-BN (Fig. 2). Continuous and reversible in-situ tunability of hybrid polaritons in graphene/h-BN meta-structures demonstrated here (Fig. 3) is a significant advantage over other artificial and natural hyperbolic materials[1-4], and is appealing from both the perspective of fundamental physics as well as potential applications[3-5,25-29]. Thus, our work uncovers a practical approach for nano-photonic meta-structures with intertwined electronic, plasmonic, phononic, and/or exciton polaritonic properties[16]. Specifically, vdW polaritonic heterostructures with locally tunable properties fulfill the essential prerequisites for the implementation of transformation two-dimensional plasmonics[30,31]. The hybridization and graphene-induced tunability reported here are expected to be generic for other electromagnetic metamaterials[32] and vdW heterostructures[8,16]. A precondition for these effects is an overlap between various polaritonic dispersion branches. Finally, we remark that it is possible to make an analogy between altering the polariton dispersion by graphene and the Goos–Hänchen effect (GHE): a lateral shift of an optical beam upon reflection from an interface[33]. Theory of such a polaritonic GHE will be reported elsewhere.

# Methods

## Experimental setup

The infrared (IR) nano-imaging and Fourier transform IR nano-spectroscopy (nano-FTIR) experiments introduced in the main text were performed using a scattering-type scanning near-field optical microscope (s-SNOM). Our s-SNOM is a commercial system (www.neaspec.com) based on a tapping-mode atomic force microscope (AFM). In the experiments, we use a commercial AFM tip (tip radius ~ 10 nm) with a $PtIr_5$ coating. The AFM tip is illuminated by monochromatic quantum cascade lasers (QCLs) (www.daylightsolutions.com), $CO_2$ lasers (www.accesslaser.com) and a broad-band laser source via difference frequency generation (DFG) (www.lasnix.com). Together, these lasers cover a frequency range of 700 – 2300 $cm^{-1}$ in the mid-IR. The s-SNOM nano-images were recorded by a pseudo-heterodyne interferometric detection module with an AFM tapping frequency 280 kHz and tapping amplitude around 70 nm. With this setup the s-SNOM is able to probe the optical signal from sub-surface objects up to a depth of ~ 250 nm. In order to subtract background signal, the s-SNOM output signal was demodulated at the $3^{rd}$ harmonics of the tapping frequency. In this work, we report our near-field data in the form of the normalized scattering amplitude using gold as the reference: $s(\omega) = s_{sample}(\omega) / s_{Au}(\omega)$.

## Sample fabrication

Hexagonal boron nitride (h-BN) crystals were mechanically exfoliated from bulk samples and deposited onto Si wafers capped with 300 nm thick $SiO_2$. Graphene was then placed onto the h-BN using a PMMA-transfer method. In this work, we use graphene from either mechanical exfoliation or chemical vapor deposition (CVD) synthesis and get similar results from both techniques. The gold film used as the reference in our measurements was lithographically fabricated on the same substrate. Electrostatic back-gating was accomplished by applying the voltage between the Si wafer and graphene layer, with $SiO_2$ and h-BN as the gate dielectrics.

**Author contributions:**

S.Z. provided CVD graphene samples used to collect data in Figs. 2 and 3. All other authors were involved in designing the research, performing the research and writing the paper.

**Competing interests:**

F.K. is one of the cofounders of Neaspec and Lasnix, producer of the s-SNOM and infrared source used in this work. All other authors declare no competing financial interests.

**Additional information:**

Supplementary information accompanies this paper at www.nature.com/naturenanotechnology.

Reprints and permission information is available online at http://npg.nature.com/reprintsandpermissions/.

Correspondence and requests for materials should be addressed to D.N.B.

**Acknowledgments:**

Work at UCSD on optical phenomena in vdW materials is supported by DOE-BES DE-FG02-00ER45799 and the Moore foundation. Research at UCSD on metamaterials and development of nano-IR instrumentation is supported by Air Force Office of Scientific Research (AFOSR), University of California Office of The President (UCOP), and Office of Naval Research (ONR). P.J-H acknowledges support from AFOSR grant number FA9550-11-1-0225.

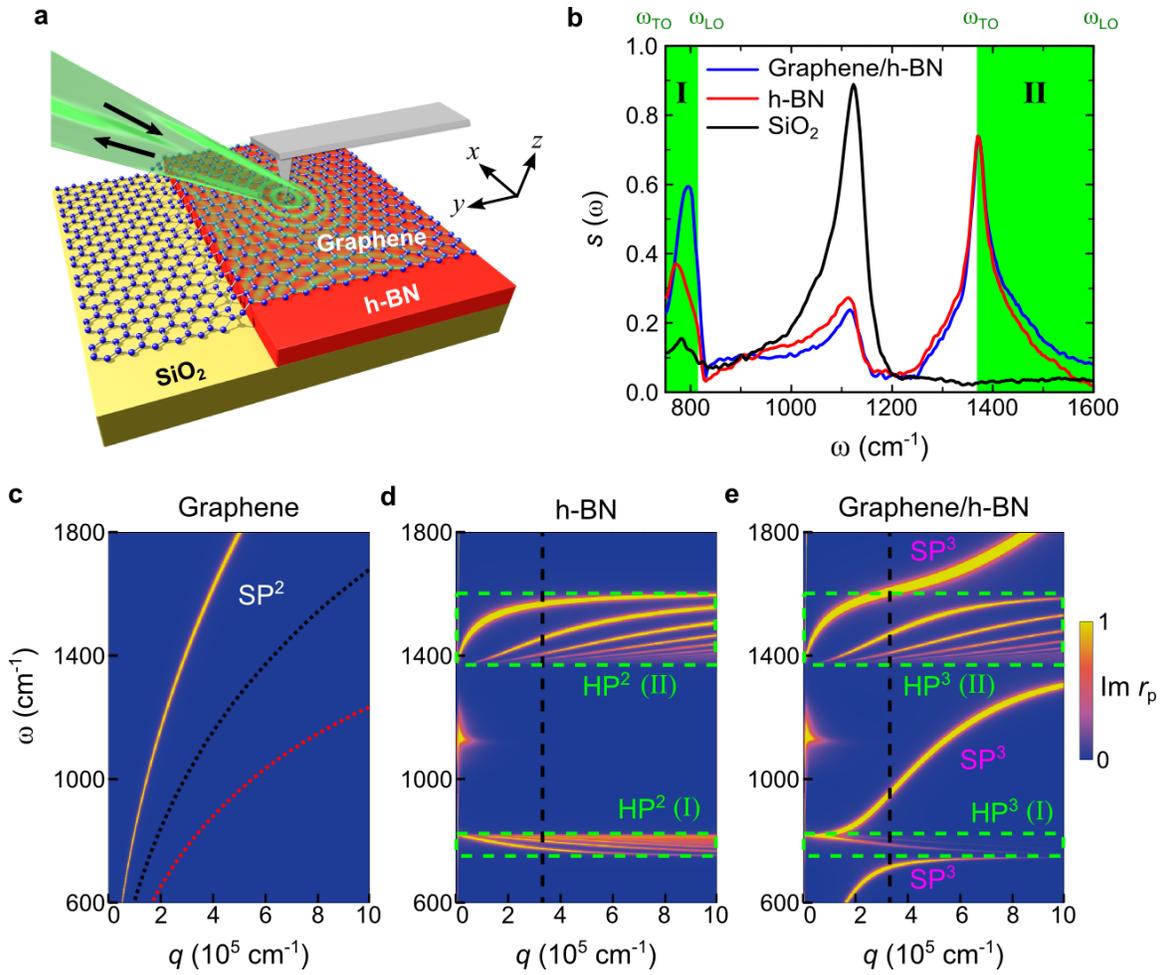

**Figure 1 | An Overview of hybridized hyperbolic response in graphene/h-BN meta-structure. a**, The experimental schematic showing the IR beams (the black arrows) incident on and back-scattered by an AFM tip. The incident beam is generated from monochromatic or broad-band laser sources (Methods). The back-scattered light is collected for extraction of the near-field signal. **b**, Broad-band nano-IR spectra of the meta-structure with a representative thickness of h-BN: 58 nm. The blue, red and black curves represent the spectra for graphene/h-BN, h-BN, and SiO$_2$, respectively. The spectra are collected far away from the sample edges where the impact of edge-reflected polaritonic waves is negligible. **c,** Calculated dispersion of the surface plasmon polariton (SP$^2$) in freestanding graphene with Fermi energy $E_F$ = 0.37, 0.15, and 0.08 eV. **d,**

Calculated dispersion of the hyperbolic phonon polariton ($HP^2$) in h-BN of thickness 58 nm. The dispersion is visualized using the false colour map of the imaginary part of the reflection coefficient $r_p$ (for the case of *P*-polarization, polarized along the *z*-axis, Supplementary Section 1). The black dashed line is a rough estimate of the momentum at which the tip-sample coupling is the strongest[20]. The green dashed rectangles surround the regions of hyperbolic response. **e,** Same as (**d**) for graphene/h-BN structure with $E_F$ = 0.37 eV. The false colour map reveals the dispersion of the hyperbolic plasmon-phonon polaritons ($HP^3$) and the surface plasmon-phonon polaritons ($SP^3$). Weak resonances around $\omega = 1130 \text{cm}^{-1}$ in (**d**) and (**e**) originate from the $SiO_2$ substrate.

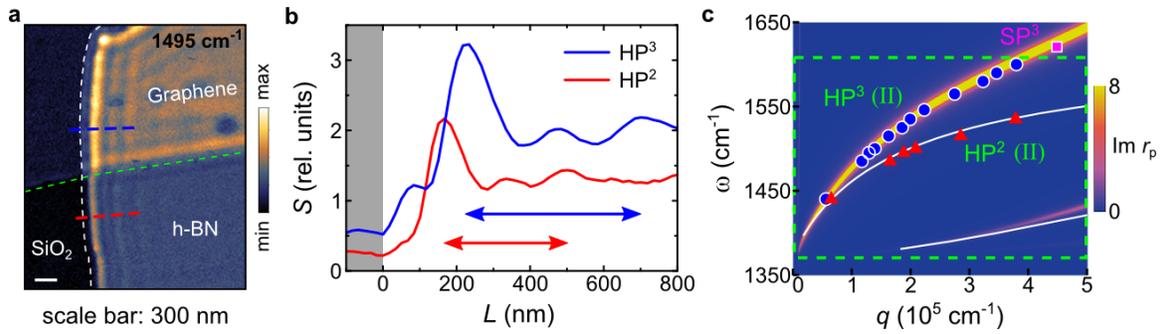

**Figure 2 | Modification of type II hyperbolic phonon polaritons in graphene/h-BN meta-structure. a**, Near-field amplitude image of the graphene/h-BN at frequency $\omega$ = 1495 cm$^{-1}$. With monolayer graphene, the intensity and wavelength of phonon polariton in pristine h-BN are increased. White and green dashed lines indicate the edge of h-BN and graphene, respectively. Scale bar: 300 nm, $d$ = 25 nm. **b**, Line profiles taken along the dashed blue and red lines in (**a**). Double arrows indicate the polariton wavelength measured on graphene/h-BN (blue) and h-BN (red). **c**, Experimental dispersion relation of type II HP$^2$ in h-BN (red triangles), HP$^3$ (blue dots) and SP$^3$ in graphene/h-BN (pink square) with the Fermi energy $E_F$ = 0.37 eV. The corresponding simulation results are also provided as the white lines and false colour map, respectively.

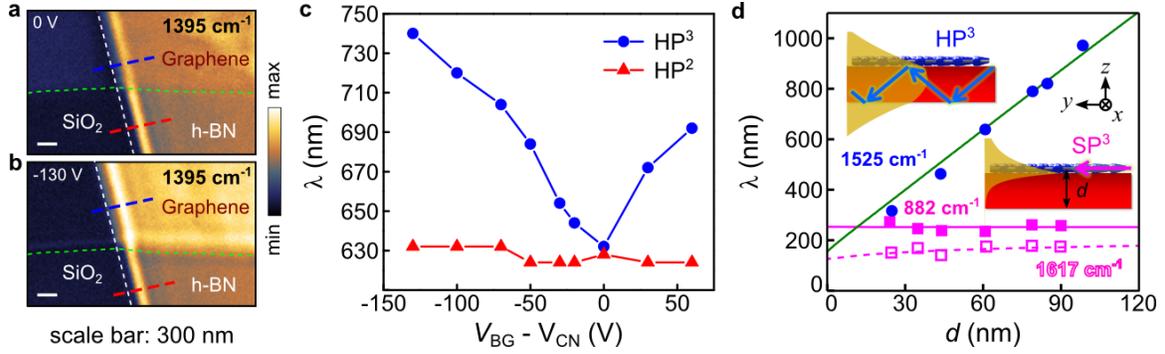

**Figure 3 | Tuning of the graphene/h-BN polariton wavelength by electrostatic gating and varying the meta-structure thickness. a-b**, Near-field images of graphene/h-BN and h-BN polaritons at back gate (BG) voltages relative to the charge neutral (CN) state $V_{BG} - V_{CN} = 0$ V (**a**) and –130 V (**b**). Scale bar: 300 nm. **c**, Gate voltage dependence of the HP$^3$ wavelength in graphene/h-BN meta-structure (blue dashed line in Figs. 3a-b) and the apparent lack of thereof for HP$^2$ in h-BN (red dashed line in Figs. 3a-b) at $\omega =$ 1395 cm$^{-1}$. Thickness of the h-BN in (**a-c**): $d = 4$ nm. **d**, The dependence of HP$^3$'s wavelength on h-BN thickness at $\omega = 1525$ cm$^{-1}$ (data and simulations are shown with the blue dots and green line, respectively). For the SP$^3$, there is no systematic thickness-dependence (solid and hollow pink squares at $\omega = 882$ and 1617 cm$^{-1}$, respectively. The corresponding simulations are plotted as the solid and dashed lines). Inset, the propagation schematics for HP$^3$ (top) and SP$^3$ (bottom). Yellow shapes in each inset show the real part of the polariton field as a function of $z$ obtained using equation (S4).